\documentclass[prc,twocolumn,superscriptaddress,nofootinbib]{revtex4-1}
\usepackage{epsfig}
\usepackage{graphicx}
\usepackage{palatino}
\usepackage[english]{babel}
\usepackage{hyphenat}
\usepackage{amsmath}
\usepackage{amssymb}
\usepackage{slashed}
\usepackage{epstopdf}
\usepackage{xcolor}
\usepackage{booktabs}
\definecolor{lcolor}{rgb}{0.,0.0,0.}
\definecolor{citcolor}{rgb}{0,0.,0.5}
\usepackage[breaklinks,colorlinks,urlcolor=blue,citecolor=blue,linkcolor=blue]{hyperref}
\usepackage{multirow}
\usepackage{ltablex}

\newcommand{\beq}{\begin{equation}}
\newcommand{\eeq}{\end{equation}}
\newcommand{\bea}{\begin{eqnarray}}
\newcommand{\eea}{\end{eqnarray}}

\newcommand{\bem}{\begin{multline}}
\newcommand{\eem}{\end{multline}}
\newcommand{\beg}{\begin{gather}}
\newcommand{\eeg}{\end{gather}}

\newcommand{\nn}{\nonumber\\}

\newcommand{\ben}{\begin{eqnarray*}}
\newcommand{\een}{\end{eqnarray*}}

\newcommand{\bal}{\begin{align}}
\newcommand{\eal}{\begin{align}}

\newcommand{\secn}[1]{Section~1}
\newcommand{\appn}[1]{Appendix~1}

\long\def\comment#1{ }

\def\med{\text{med}}

\def\and{\quad\text{and}\quad}

\def\cut{\text{cut}}

\def\0{{\boldsymbol 0}}

\begin{document}
\title{{\bf A background estimator for jet studies in p+p and A+A collisions\\}}
\author{Yacine Mehtar-Tani}
\email[]{mehtartani@bnl.gov}
\affiliation{Physics Department, Brookhaven National Laboratory, Upton, NY 11973, USA.}
\author{Alba Soto-Ontoso}
\email[]{ontoso@bnl.gov}
\affiliation{Physics Department, Brookhaven National Laboratory, Upton, NY 11973, USA.}
\author{Marta Verweij}
\email[]{m.verweij@uu.nl}
\affiliation{Department of Physics and Astronomy, Vanderbilt University, Nashville, TN 37235, USA.}
\affiliation{RIKEN BNL Research Center, Brookhaven National Laboratory, Upton, NY 11973, USA.}
\affiliation{Institute for Subatomic Physics, Utrecht University, Utrecht, Netherlands.}

\begin{abstract}
Experimentally, jet physics studies face an unavoidable task: distinguishing, at the detector level, the particles produced in the hard partonic scattering from the ones created in  unrelated soft processes such as pileup interactions in high-luminosity proton-proton scattering or the underlying event in heavy-ion collisions. The fluctuating nature of the background constitutes the main source of uncertainty for  any subtraction algorithm. Aiming at mitigating the effect of such fluctuations, we present a new method to estimate the background contribution to the transverse momentum on a jet-by-jet basis. Our approach is based on estimating the median background momentum density stored above a $p_{\rm T}$-cut applied at the constituent level and an experimentally accessible correction term related to the signal contribution below the cut. This allows to trade part of the uncertainty due to background contamination for that of the signal below the cut, similarly to SoftKiller method. We propose to reduce the fluctuations of the latter by exploiting intrinsic correlations among the soft and hard sectors generated in the branching process of QCD jets. Our data-driven approach is tested against PYTHIA8 and JEWEL di-jet events embedded in  minimum bias events and thermal background, respectively, and compared to the area-median and SoftKiller methods. The main result of this study is a $\sim 5\!-\!35\%$ improvement on the resolution of the reconstructed jet $p_T$ compared to previous methods in both high-luminosity proton-proton and heavy-ion collisions. 
\end{abstract}

\maketitle
\section{Introduction}
The study of jets in ultra-relativistic hadronic collisions has decisively contributed to our understanding of the perturbative and non-perturbative aspects of Quantum Chromodynamics (QCD).
Currently, at the Large Hadron Collider (LHC), events with jets in the final state are being extensively exploited to search for signatures of new physics. In the context of heavy-ion collisions, jet physics is immersed into the precision era both from a theoretical and an experimental perspective. Traditional observables, sensitive to the formation of a Quark-Gluon Plasma (QGP), such as the nuclear modification factor~\cite{Adcox:2001jp,Aamodt:2010jd} or di-jet asymmetry~\cite{Aad:2010bu}
, are being complemented with substructure analysis techniques inherited from the p+p community both at RHIC and the LHC~\cite{Sirunyan:2017bsd, Kauder:2017mhg, Andrews:2018jcm}.

To perform meaningful theory-to-data comparisons, jet physics relies on two main ingredients: jet reconstruction algorithms and background subtraction methods. The former provide an operational definition of a jet by clustering the particles in the event following an infrared and collinear-safe method based on the notion of a distance between them either in momentum ($k_T$) or angular space~\cite{Catani:1993hr,Dokshitzer:1997in,Cacciari:2008gp}. The latter deals with the fact that the situation in which an isolated hard partonic scattering is responsible for all the clustered particles is hardly realized in current collider experiments. In high-energy hadronic collisions two competing mechanisms are relevant: pileup interactions, ubiquitous in high-luminosity facilities and the underlying event whose importance grows with the collision system size, i.e. from p+p to A+A~\cite{Abelev:2012ej}. Both of them populate the low-$p_T$ sector of the event and are commonly referred to as \textit{background}. The aforementioned clustering algorithms do not distinguish signal from background particles. Consequently, any reconstructed jet property $X$ has two independent contributions, i.e. $X^{\rm reco}_{\rm jet}\!=\!X^{\rm sig}_{\rm jet}+X^{\rm bkg}_{\rm jet}$. Background subtraction methods are developed to subtract an estimated value for the $X^{\rm bkg}_{\rm jet}$ term. We shall focus in the present work on jet transverse momentum reconstruction so that $X\!=\!p_T$. To this aim, two steps have to be addressed. First, the value of $p^{\rm bkg}_{T,\rm jet}$ has to be estimated. This is an intricate task due to the fact that the background contribution fluctuates on an event-by-event basis and locally within the same event. Next, the question on how to remove this estimated background at the particle level arises. These two elements are strongly correlated as an improvement on the $p^{\rm bkg}_{T,\rm jet}$ estimate largely impacts the reconstruction of other features such as the jet mass~\cite{Acharya:2017goa, ATLAS:2018jsv, sharedLayer}.  In what follows we focus on background estimation. The subtraction part will be addressed in a future work. 
 
 Several methods are used to estimate the background transverse momentum contribution on a jet-by-jet basis in p+p and A+A collisions~\cite{Cacciari:2007fd,Cacciari:2014gra,Berta:2014eza, Haake:2018hqn}. In this work we would like to draw the attention on the subtle interplay between signal and background in the low-$p_T$, few GeV, region. We pinpoint the underlying mechanisms driving the mean and standard deviation of the reconstructed jet momentum distribution. This systematic study leads to the design of a new background estimator that improves the jet $p_T$ reconstruction resolution. Furthermore, we propose an experimentally measurable quantity, to be described in the following, to correct for signal contamination. Our method, dubbed $\rho$-correction as it corrects for the background energy density, aims at mitigating both background and signal fluctuations on a jet-by-jet basis.

This paper is organized as follows. In the next Section we provide a detailed explanation of the generator samples together with a review of existing background estimators and emphasize the importance of signal contamination. The new approach and its application to p+p and A+A collisions are presented in Section~\ref{Section3}. To conclude, future lines of work are outlined in Section~\ref{Section5}.

\section{Background estimation}
In this Section, the simulation framework together with the key ingredients of existing methods that aim to provide an accurate estimate of the background transverse momentum contribution to a jet $p_T$ are exposed. In the last part of this Section we analyze the impact of signal contamination below a given soft $p_{T}^{\cut}$.
\subsection{Simulation set-up}
First, we introduce the toy data samples that have been used to compute all the results presented in this manuscript. 

The situation of a p+p collision at 13 TeV with large pileup is mimicked by superimposing a PYTHIA8~\cite{Sjostrand:2007gs} di-jet event into a number ${\rm{n_{PU}}}$ of minimum bias events. The virtuality of the hard partonic scattering is set to $\hat p_T\!=\!100$~GeV/c and the number of pileup interactions is set to ${\rm{n_{PU}}}\!=\!200$ for the high-luminosity LHC~\cite{Atlas:2019qfx}. For completeness, we also consider current LHC running conditions with ${\rm{n_{PU}}}\!=\!60$.

In the case of a central heavy-ion collision, which will be addressed in Sec.~\ref{sec:hic}, we make use of the JEWEL~\cite{Zapp:2011ya} event generator without recoil (JEWEL NR) embedded into a thermal background, as a proxy for the QGP. Di-jet events are produced at a center-of-mass energy per nucleus-nucleus collisions $\sqrt s\!=\!5.02$~TeV. Then, the parton shower is modified by collisions with the background constituents and medium-induced radiative processes. Background particles are obtained by sampling a Boltzmann distribution with a fixed multiplicity $\langle N \rangle\!=\!7000$ and an average momentum $\mu\!=\!1.2$~GeV/c. This corresponds to an average momentum density of $\langle\rho\rangle\!=\!250$~GeV, a realistic value for central Pb+Pb collisions. In this case, background particles are massless and uniformly distributed in ($\eta$, $\phi$)-space with $|\eta|\!<\!3$.

For each event, background and signal particles are clustered into anti-$k_T$~\cite{Cacciari:2008gp} jets with $R\!=\! 0.4$ using FastJet3.1~\cite{Cacciari:2011ma}. The true, $p_{T,\rm{truth}}^{\rm jet}$, and the reconstructed, $p_{T,\rm{reco}}^{\rm jet}$, jet momenta are defined as follows
\bea
&p_{T,\rm{truth}}^{\rm jet}=\displaystyle\sum_i p_{T,i\in{\rm sig}} \\
&p_{T,\rm{reco}}^{\rm jet}=p_{T,\rm{raw}}^{\rm jet}-p_{T,\rm{bkg}}^{\rm jet}
\eea
where $p_{T,\rm{raw}}^{\rm jet}$ refers to the jet momentum resulting from the clustering and the subscript ${\rm sig}$ labels the particles from PYTHIA/JEWEL. The goal of this paper is to provide an estimate of $p_{T,\rm{bkg}}^{\rm jet}$ that leads to a ($p_{T,\rm{reco}}^{\rm jet}\!-\!p_{T,\rm{truth}}^{\rm jet}$)-distribution with a close to zero mean and minimal standard deviation.
Finally, the analysis is performed on jets with $p_{T,\rm{truth}}^{\rm jet}\!>\!120$~GeV/c. 

\subsection{Previous methods}\label{Section2}

The most largely used background estimation method by the jet experimental community is the area-median~\cite{Cacciari:2007fd} whose biggest strength is its data-driven and unbiased nature. On an event-by-event basis, the particles are clustered with the $k_T$ algorithm~\cite{Catani:1993hr}. After discarding the two hardest clusters (patches), under the assumption that they are predominantly composed of signal particles, $\rho$ defined as 
\beq
\rho=\displaystyle\frac{p_{T}}{A}
\eeq
is computed for each cluster, where $p_{T}$ and $A$ are the total transverse momentum and area in the $(\eta,\phi)$ plane of the corresponding patch, respectively. The background contribution to the reconstructed jet $p_T$ is then given by
\beq
p_{T,\rm{bkg}}^{\rm{jet}}\Big\vert_{\rm area-median} = {\rm{med}(\rho)}A,
\eeq
where ${\rm{med}(\rho)}$ is the median $p_T$ density of the set of patches. 

Although highly successful at estimating $p_{T,\rm{bkg}}^{\rm{jet}}$ on average, this method lacks control over background fluctuations. This translates into a ($p_{T,\rm{reco}}^{\rm jet}\!-\!p_{T,\rm{truth}}^{\rm jet}$)-distribution centered at zero but, as shown in Section~\ref{Section3}, with a sizable standard deviation~\cite{Abelev:2012ej}. 

An effective way of improving the resolution of the area-median method is to introduce a $p_T^\cut$ at the constituent level. If signal and background do not overlap in momentum space, removing all particles in the event below the separation scale would result into an exact estimation of the background in each patch (jet). In this ideal scenario the ($p_{T,\rm{reco}}^{\rm jet}\!-\!p_{T,\rm{truth}}^{\rm jet}$)-distribution would be a Dirac Delta. This is the underlying idea behind SoftKiller (SK)~\cite{Cacciari:2014gra}. 

In this case, the event is clustered using a grid in ($\eta,\phi$)-space of size \textit a that is a free parameter of the method whose role is discussed below~\footnote{In this work we use $k_T$-clustering instead of a rectangular grid to ensure an apples-to-apples comparison with the $\rho$-correction approach in what follows.}. For every cell in the grid, the maximum constituent's momentum, $p_{T,i}^{\rm max}$, is found in order to compute the $p_T^\cut$ defined as 
\beq
p_{T}^{\cut,\rm SK}={\rm{med}}(p_{T,i}^{\rm max}).
\label{SKcut}
\eeq
Then, the jet-by-jet background estimate of this method is expressed like
\beq
p_{T,\rm{bkg}}^{\rm{jet}}\Big\vert_{\rm SK} =\displaystyle\sum^{p_{T,i} <p_{T}^{\cut,\rm SK}}_i p_{T,i}.
\eeq
Behind its apparent simplicity, this algorithm hides a few remarkable subtleties that we would like to comment on. Clearly, SoftKiller's performance depends on the scale separation between background and signal. In general, soft signal particles would be removed if they fall below the $p_T^\cut$ leading to a narrower but shifted to negative values ($p_{T,\rm{reco}}^{\rm jet}\!-\!p_{T,\rm{truth}}^{\rm jet}$)-distribution. An effective solution to overcome this bias is to tune the size of the patches where the $p_T^{\rm{cut}}$ is computed. That is, by construction there is a value of the parameter $a$ or equivalently $R_{kT}$ (the resolution parameter in the $k_T$ clustering) for which the shift of the ($p_{T,\rm{reco}}^{\rm jet}\!-\!p_{T,\rm{truth}}^{\rm jet}$)-distribution is zero. This is so because a non-trivial interplay exists between the grid size/clustering radius and the $p_{T}^{\cut}$: modifying $R_{kT}$ is equivalent to change the value of $p_{T}^{\cut}$ and, consequently, the grid size can be adjusted to balance the signal contamination and background contribution below and above the cut. Therefore, in the following, we tune the value of $R_{kT}$ such that the absolute mean value of the $p_T$ reconstruction distribution is smaller than $0.5$~GeV/c. We shall refer to this method as "SoftKiller $R^{\rm{adj}}$". Notice that this choice does not constrain, in principle, the standard deviation of the momentum reconstruction distribution. Finally, we would like to comment on the fact that, in this work, we use SoftKiller as a background estimator and not as a subtraction method. In our opinion, the resulting $p_{T,\rm{bkg}}^{\rm{jet}}$ could be used as an input to different subtraction algorithms, e.g. \cite{Berta:2014eza}, to avoid distortions in the jet substructure due to the removal of all particles below $p_{T}^{\cut}$.  
\subsection{Impact of signal contamination}
\begin{figure}[ht]
\begin{center}
\includegraphics[scale=0.4]{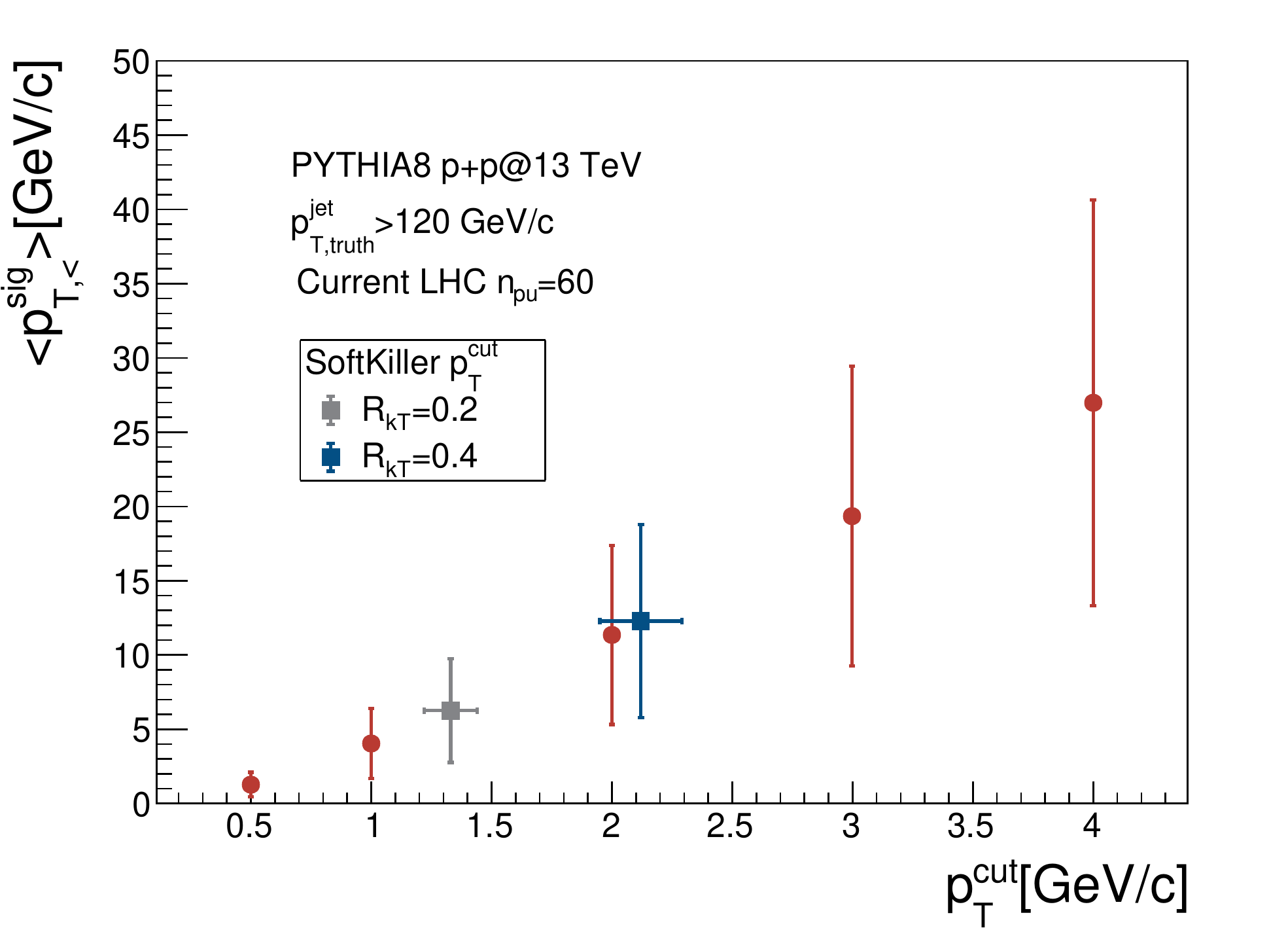} 
\end{center}
\vspace*{-0.5cm}
\caption[a]{Average $p_T$ contribution from signal particles (see Eq.~\ref{signalCorrection}) as a function of $p_T^{\cut}$ (red squares). The SoftKiller cut as given by Eq.~\ref{SKcut} is represented by gray ($R_{kT}\!=\!0.2$) and blue ($R_{kT}\!=\!0.4$) squares. Error bars reflect the standard deviation.}
\label{fig1}
\end{figure}

In the previous Section, we remarked that any background estimator that relies on a soft momentum cut is sensitive to QCD radiation below this threshold. To quantify the signal contamination, we have computed the average contribution of the total jet $p_T$ accounted by constituents whose momentum is below a certain $p_T$ threshold:
\beq
 p_{T,<}^{\rm sig} = \displaystyle\sum^{p_{T,i} <p_{T}^{\cut}}_{i\in{\rm sig}} p_{T,i}.
\label{signalCorrection}
\eeq
 The event-averaged value of this quantity, for the PYTHIA8 sample, as a function of multiples of the momentum cut at the constituent level, $p_{T}^{\cut}$, is shown in Fig.~\ref{fig1}. For completeness, we include the values of SoftKiller $p_T^{\rm cut}$ obtained by clustering the event containing both signal and background particles from ${\rm{n_{PU}}}\!=\!60$ minimum bias events into $k_T$-patches of $R_{kT}\!=\!0.2,0.4$, removing the two hardest ones and then applying Eq.~\ref{SKcut}. As expected, the signal contribution is a monotonous increasing function of $p_T^\cut$. It is worthwhile to mention that even for considerably low thresholds such as $2$~GeV/c the contribution of signal particles amounts to almost 15 GeV/c, which is more than $10\%$ of the jet $p_T$ selection threshold. This indicates a strong overlap between signal and background. Further, the aforementioned direct relationship between $R$ and the SoftKiller $p_{\rm T}^{\cut}$ (see Eq.~\ref{SKcut}) is explicitly shown, i.e. $p_{\rm T}^{\cut}$ grows when increasing $R_{kT}$. More concretely, increasing the $p_{\rm T}^{\cut}$ by $1$~GeV leads to a signal contamination growth of $\sim 5$~GeV. This is an important aspect given the fact that SoftKiller $R^{\rm{adj}}$ is based on canceling out the signal contamination with the background contribution above the cut. That is, minimizing the shift requires a precise fine-tuning of $R_{kT}$ and small deviations from the optimal value would result into a sizable bias in the $p_T$ reconstruction.

\begin{figure}[ht]
\begin{center}
\includegraphics[scale=0.4]{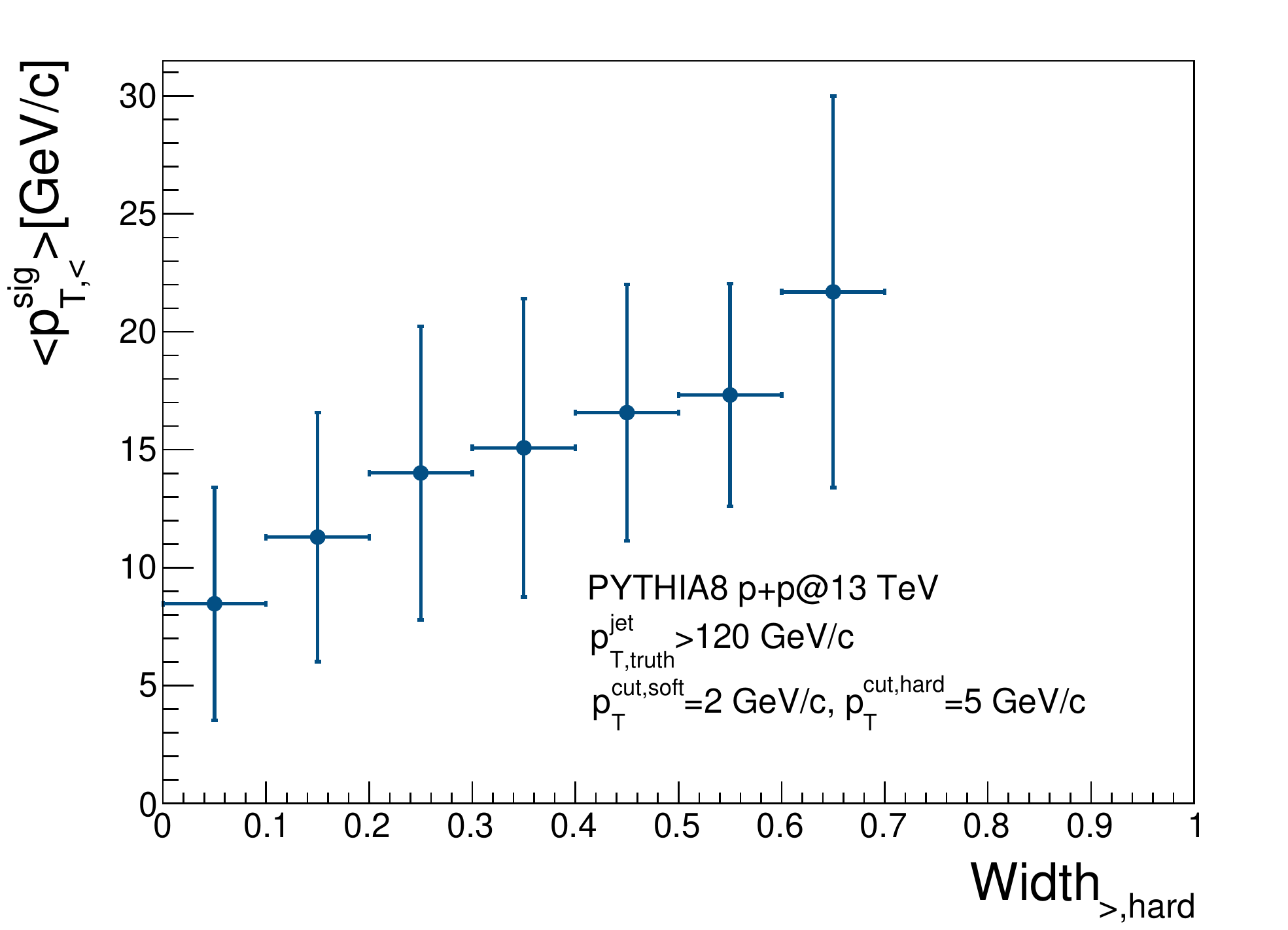}  
\end{center}
\vspace*{-0.5cm}
\caption[a]{Average $p_T$ contribution from signal particles (see Eq.~\ref{signalCorrection}) as a function of the width above the hard cut (see Eq.~\ref{width}). The vertical error bars reflect the standard deviation while the horizontal ones represent the bin size.}
\label{fig2}
\end{figure}

In Fig.~\ref{fig1} we have presented an inclusive computation of $\langle p_{T,<}^{\rm{sig}}\rangle$, i.e. integrating out the information about other jet features. However, QCD dynamics generates intrinsic correlations between the soft and hard $p_T$ sectors within a jet owing to the tree structure of QCD parton cascades. The constituents whose momenta is below a given $p_T^{\cut,{\rm soft}}$, responsible for the $\langle p_{T,<}^{\rm{sig}}\rangle$-term, originate from the branching of hard $p_T$ partons. Each of these multiple soft radiations is not independent from the rest of the shower as its phase space is restricted by mechanisms such as energy conservation or color coherence. This correlated radiation pattern is expected to manifest itself in jet substructure observables. For instance, an enhancement of soft radiation for broader jets is expected. We test this hypothesis within PYTHIA8 and the result is shown in Fig.~\ref{fig2} where the width is defined as~\cite{Larkoski:2014pca}
\beq
{\rm{width}_{>,{\rm{hard}}}}=\displaystyle\sum^{p_{T,i} >p_T^{\cut,{\rm hard}}}_i \displaystyle \frac{p_{T,i}}{p_{T,\rm {jet}}}\displaystyle \frac{\Delta R_{i}}{R_{\rm {jet}}}
\label{width}
\eeq
and $p_T^{\cut,{\rm hard}}$ is a hard momentum threshold above which the width is computed, $R$ refers to the jet radius and $\Delta R_i$ is the cylinder ($\eta,\phi$)-distance from each constituent to the jet axis. A clear correlation between $\langle p_{T,<}^{\rm{sig}}\rangle$ and ${\rm{width}_{>,{\rm{hard}}}}$ is observed. Further, the standard deviation of $\langle p_{T,<}^{\rm{sig}}\rangle$ is reduced when fixing ${\rm{width}_{>,{\rm{hard}}}}$ as compared to the inclusive case (see Fig.~\ref{fig1}). Therefore, signal fluctuations in the low-$p_T$ regime can be mitigated by using hard-soft correlations within a jet. For simplicity, we restrict the results of this manuscript to the width and leave a more systematic study of other metrics and combinations of them for future work.

All the elements discussed in this Section, i.e. a momentum cut, the median $\rho$ and the average signal in the soft sector, are combined into a new background estimator to be described below.
\section{The $\rho$-correction method}\label{Section3}
On an event-by-event basis, the $\rho$-correction algorithm proceeds as follows:
\begin{enumerate}
\item Divide the event into patches with $R_{kT}\!=\!0.4$ via the $k_T$ algorithm  and remove the two hardest ones.
\item For each patch, compute $\rho$ with the particles whose $p_T$ is above a $p_T^{\cut,{\rm soft}}$ (which will be discussed below)
\beq
\rho_>=\displaystyle\frac{1}{A}\displaystyle\sum^{p_{T,i} >p_T^{\cut,{\rm soft}}}_i p_{T,i}
\label{ptabove}
\eeq
where $A$ denotes the area of the patch.
\item Find the median of these $\rho$ values,  $\med(\rho_>)$. 
\item Re-cluster the whole event via the anti-$k_T$ algorithm with $R\!=\!0.4$.
\item For each anti-$k_T$ cluster, add up the $p_T$'s of those particles that are below $p_T^{\cut,{\rm soft}}$: 
\beq
p_{T,<}=\displaystyle\sum^{p_{T,i} <p_T^{\cut,{\rm soft}}}_i p_{T,i}
\label{ptbelow}
\eeq
\item The final estimate for the background $p_T$ of a given anti-$k_T$ jet is 
\beq
p_{T,\rm{bkg}}^{\rm{jet}}\Big\vert_{\rho-{\rm correction}}= \med(\rho_>)A + p_{T,<} -  p_{T,<}^{\rm{shift}}
\label{ourCut}
\eeq 
where the subscripts "$<$" , "$>$" refer to the region where these quantities are computed, i.e. below or above the $p_T^{\cut,{\rm soft}}$.
\end{enumerate}
The pocket-formula of the $\rho$-correction method, Eq.~\ref{ourCut} neglecting the last term, interpolates between the area-median and SoftKiller for which  $p_T^{\cut,{\rm soft}}\!=\!0$ and $p_{T}^{\cut,\rm SK}$ (see Eq.~\ref{SKcut}), respectively. 

The role of the last element, $p_{T,<}^{\rm{shift}}$, is to correct for the offset induced by signal contamination (see Figs.~\ref{fig1},\ref{fig2}). To address the definition of $p_{T,<}^{\rm{shift}}$, an explicit distinction between jets arising from p+p and A+A collisions is needed. This is due to the impact of the well established phenomenon of energy loss i.e. parton showers are modified in the presence of a QGP leading to a degradation on the jet $p_T$ with respect to its parent parton~\cite{Mehtar-Tani:2013pia}.

\begin{itemize}
    \item {{\bf{p+p}}: we explore two possibilities for $p_{T,<}^{\rm{shift}}$. First, using an inclusive determination of $\langle p_{T,<}^{\rm sig} \rangle$ ($\rho$-correction) i.e. $p_{T,<}^{\rm{shift}}\!=\!\langle p_{T,<}^{\rm sig} \rangle$. Second, the background $p_T$ estimate of each jet is corrected for signal contamination with a fluctuating value of $p_{T,<}^{\rm{sig}}$ depending on its ${\rm{width}_{>,{\rm{hard}}}}$ such that 
    \beq
    p_{T,<}^{\rm{shift}} = (p_{T,<}^{\rm sig}\big|{\rm{width}_{>,{\rm{hard}}}})
    \eeq
    Given the fact that we exploit hard-soft correlations we name this option "$\rho$-correction HS$_{\rm{corr}}$". This choice requires a value of $p_T^{\cut,{\rm{hard}}}$ so that the width is computed in the background-free region. We fix it to be $5\mu$. We have checked that the results are robust against modifications of this value. Also, when $p_T^{\cut,{\rm{hard}}}\gg 5\mu$ the soft and hard sectors become de-correlated and the inclusive result is recovered. At this point the natural question on how to extract this information, outside a Monte Carlo environment, arises. We propose to measure it at low pileup proton-proton collisions. The idea of using low pileup events as relatively background-free samples was also explored in ~\cite{Komiske:2017ubm}.
We have studied the behavior of $\langle p_{T,<}^{\rm sig} \rangle$ with the jet $p_T$ finding a logarithmic dependence. This is an important aspect in terms of applicability of the method to experimental data as inaccuracies in the determination of the true jet $p_T$ due to detector effects or background contamination would not substantially affect the value of $\langle p_{T,<}^{\rm sig}\rangle$. Therefore, the $\rho$-correction method ensures a ($p_{T,\rm{reco}}^{\rm jet}\!-\!p_{T,\rm{truth}}^{\rm jet}$)-distribution that is centered around zero thanks to the $p_{T,<}^{\rm{shift}}$-term.}
    \item {{\bf{A+A}}: the impossibility of experimentally accessing a background-free environment in a heavy-ion collision in which $p_{T,<}^{\rm{shift}}$ could be determined in an unbiased way demands an alternative definition of it. For the $\rho$-correction case we propose 
     \beq
    p_{T,<}^{\rm{shift}} = \langle p_{T,<} \rangle - \langle \rm{med}(\rho_{<})A) \rangle,
    \label{shift_aa}
    \eeq
    where $\langle p_{T,<} \rangle$ is obtained by performing an ensemble average of Eq.~\ref{ptbelow} over a set of unsubtracted jets i.e. containing signal and background particles. The second term in Eq.~\ref{shift_aa}, $\rm{med}(\rho_{<})$, is extracted from the same sample of $k_T$ jets that we used to compute Eq.~\ref{ptabove}. Again, $\langle \cdot\rangle$ represents an ensemble average over the unsubtracted jets with area $A$. Remarkably, every quantity in Eq.~\ref{shift_aa} can be extracted from data making the proposed method directly applicable to experimental analyses in heavy-ion collisions. If hard-soft correlations where to be exploited, the correction reads as follows
    \begin{align}
    \label{shift_aa_hscorr}
    p_{T,<}^{\rm{shift}} &= (p_{T,<}^{\rm sig}\big|{\rm{width}_{>,{\rm{hard}}}})_{{\rm{p+p}}} + \langle p_{T,<} \rangle  \nn 
    &  - \langle {\rm{med}}(\rho_{<})A) \rangle -\langle p_{T,<}^{\rm sig} \rangle_{{\rm{p+p}}}.
    \end{align}
    The two extra terms in Eq.~\ref{shift_aa_hscorr}, as compared to Eq.~\ref{shift_aa}, have a transparent interpretation. As previously discussed, $(p_{T,<}^{\rm sig}\big|{\rm{width}_{>,{\rm{hard}}}})_{{\rm{p+p}}}$ cannot be determined from heavy-ion data. Therefore, one has to use the same correction as in the p+p case i.e. obtained from low pileup events. Clearly, this introduces an additional shift that can be compensated with $\langle p_{T,<}^{\rm sig} \rangle_{{\rm{p+p}}}$, defined in Eq.~\ref{signalCorrection}.
    }
\end{itemize}

Finally, the determination of $p_T^{\cut,{\rm soft}}$ proceeds as follows. We select a fixed value of $p_T^{\cut,{\rm soft}}$ requiring that it minimizes the standard deviation of the ($p_{T,\rm{reco}}^{\rm jet}\!-\!p_{T,\rm{truth}}^{\rm jet}$)-distribution. We have checked that the optimal value of $p_T^{\cut,{\rm soft}}$ does not strongly depend on whether the inclusive or exclusive version of $\langle p_{T,<}^{\rm sig} \rangle$ is used to correct for the signal contamination. Hence, we believe that the most direct way to apply our method to experimental data is to use Eq.~\ref{ourCut} with the inclusive value of $\langle p_{T,<}^{\rm sig} \rangle$ to find the optimal $p_T^{\cut,{\rm soft}}$. Once this value is known one can use the width-dependent version of $\langle p_{T,<}^{\rm sig} \rangle$ to get a further reduction on the $p_T$ resolution. 
The results of the method in two different contexts are presented in the next Section.

\subsection{Proton-proton collisions}\label{pp}
\begin{figure}[ht]
\begin{center}
\includegraphics[scale=0.4]{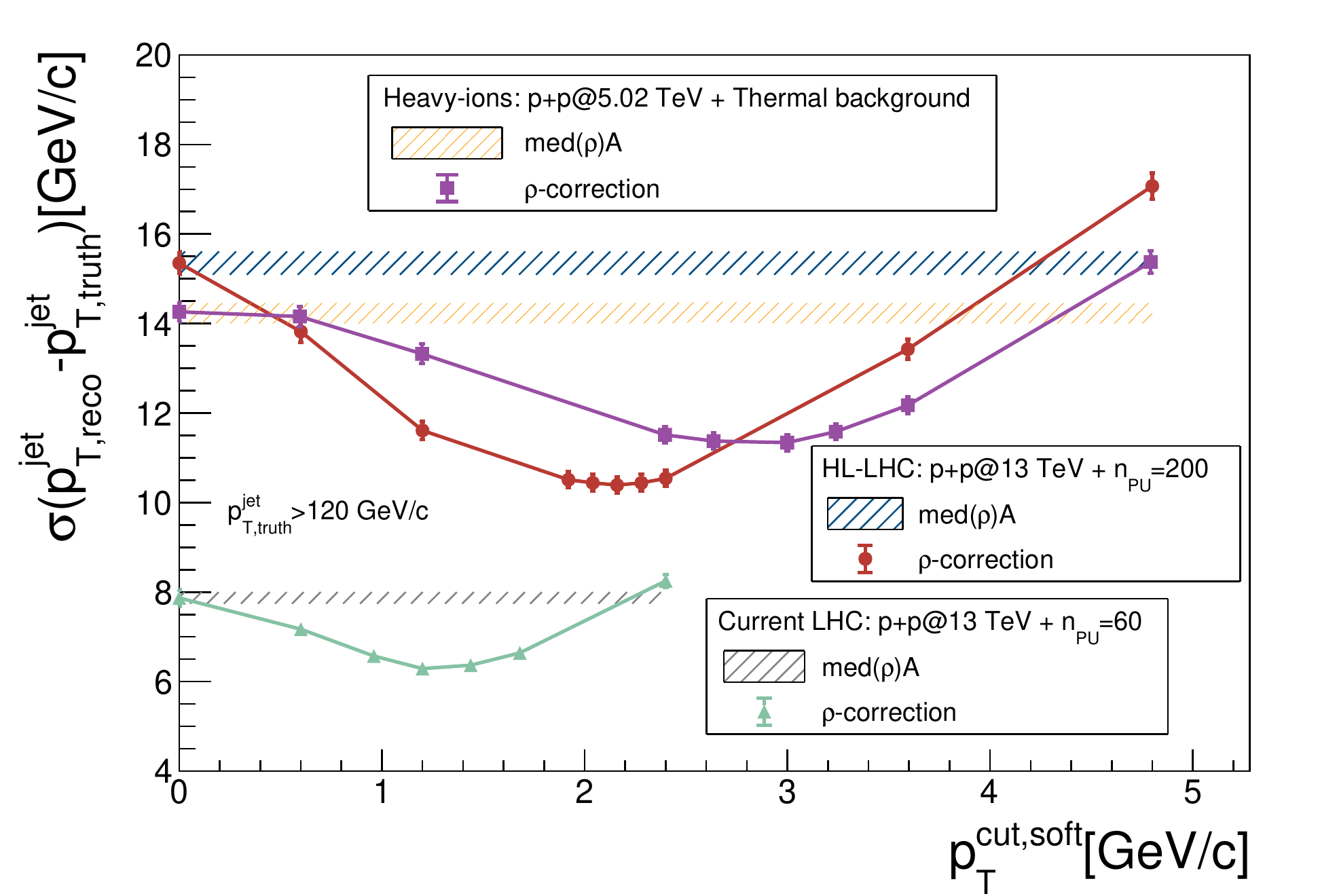}
\end{center}
\vspace*{-0.5cm}
\caption[a]{Standard deviation for the jet $p_T$ reconstruction distributions as a function of $p_T^{\cut,{\rm{soft}}}$ for the area-median approach and $\rho$-correction for three different scenarios: HL-LHC ($\rm{n_{PU}\!=\!200}$) (dashed blue, red dots), ($\rm{n_{PU}\!=\!60}$) (dashed grey, green triangles), JEWEL vacuum embedded in thermal background ($\langle\rm{N}\rangle\!=\!7000, \mu=1.2 GeV/c$) (dashed orange, purple squares). Error bars reflect statistical fluctuations.}
\label{fig3}
\end{figure}

We first test the performance of the method on p+p collisions with different pileup settings: ${\rm{n_{PU}}}\!=\!60$, the actual running conditions of the LHC and ${\rm{n_{PU}}}\!=\!200$, the foreseen scenario at the high-luminosity phase. In Fig.~\ref{fig3}, the resolution of the reconstructed jet $p_T$ as a function of $p_T^{\cut,{\rm soft}}$ for both cases is displayed. The generic shape of this curve can be understood using Eq.~\ref{ourCut}. For low values of $p_T^{\cut,{\rm soft}}$, $\rm{med}(\rho_>)$ dominates in Eq.~\ref{ourCut} and, as expected, no significant improvement is achieved with respect to the traditional area-median method. In the opposite regime, when $p_T^{\cut,{\rm soft}}\!>\!p_{T}^{\cut,\rm SK}$, the contribution of $\rm{med}(\rho_>)$ is identically zero and $\sigma$ monotonically increases with $p_T^{\cut,{\rm{soft}}}$ due to an increased number of signal particles being misidentify as background, as shown in Fig.~\ref{fig1}. 

The minimal resolution is achieved for $p_T^{\cut,{\rm {soft}}}\!=\!1.4$~GeV/c in the ${\rm{n_{PU}}}\!=\!60$ case while in the high-luminosity scenario the $p_T$-resolution is optimized for $p_T^{\cut,{\rm {soft}}}\!=\!2.2$~GeV/c. In the case of SoftKiller, the optimal values of $R_{kT}$ are $R_{kT}\!=\!0.23$ for ${\rm{n_{PU}}}\!=\!60$ and $R_{kT}\!=\!0.22$ for ${\rm{n_{PU}}}\!=\!200$. We observe a weak dependence of the standard deviation on $p_T^{\cut,{\rm soft}}$ in the region near the minimum as shown in Fig.~\ref{fig3}. This is a clear advantage with respect to SoftKiller where a precise fine-tuning was needed in order to minimize the shift. 

The precise values for the mean and standard deviation of the $p_T$-reconstruction distributions for both background settings can be found in Tables~\ref{table2},\ref{table3} for the four explored methods: area-median, SoftKiller $R^{\rm{adj}}$,$\rho$-correction and $\rho$-correction HS$_{\rm{corr}}$ where the width-dependent $\langle p_{T,<}^{\rm{sig}} \rangle$ correction is obtained for the optimal value of $p_T^{\cut,{\rm soft}}$. First of all, any method that includes a momentum threshold highly reduces the standard deviation as compared to the area-median. We find that the $p_T$ resolution of $\rho$-correction and SoftKiller $R^{\rm{adj}}$ are comparable. These two methods, although approaching the problem from different angles, are in fact similar and, as noted in \cite{Cacciari:2014gra}, the value of $R_{kT}$ for which the shift is minimized results into a close-to-minimal resolution. In the high-luminosity scenario, the $\rho$-correction HS$_{{\rm corr}}$ method results into a (35$\%$, 5$\%$) improvement on the standard deviation with respect to area-median and SoftKiller, respectively. Remarkably, our method is resilient against the particular density of the background i.e. the general trends persist when reducing the number of minimum bias events as observed in the bottom part of Fig.~\ref{fig3}. As in the previous case, the best performance is achieved by $\rho$-correction HS$_{{\rm corr}}$, i.e. when $\langle p_{T,<}^{\rm{sig}} \rangle$ is computed for bins of ${\rm{width}_{>,{\rm{hard}}}}$. We conclude that the use of a fundamental property of QCD such as hard-soft correlations improves the background momentum estimation in jet experimental analyses. This is one of the main results of this paper. 

\begin{table}[ht]
\centering
\begin{tabular}{|c||c|c|c}
\hline
 [GeV/c]&Mean& Standard deviation \\
\hline\hline
med($\rho$)A &-0.87$\pm$0.37 & 15.35$\pm$0.26 \\
SoftKiller $R^{\rm{adj}}$ &-0.29$\pm$0.25 & 10.37$\pm$0.18 \\
$\rho$-correction & 1.20$\pm$0.25&  10.39$\pm$0.18 \\
$\rho$-correction HS$_{\rm{corr}}$ &1.26$\pm$0.25& 10.31$\pm$0.18 \\
\hline
\end{tabular}
\caption{Mean and standard deviation of the $p_T$ signal reconstruction distributions for PYTHIA8 events embedded into $\rm{n_{PU}\!=\!200}$ minimum bias.}\label{table1}
\end{table}

\begin{table}[ht]
\centering
\begin{tabular}{|c||c|c|c}
\hline
 [GeV/c]&Mean& Standard deviation \\
\hline\hline
med($\rho$)A &-1.18$\pm$0.19 & 7.87$\pm$0.13 \\
SoftKiller $R^{\rm{adj}}$ & -0.49$\pm$0.15 & 6.36$\pm$0.11 \\
$\rho$-correction & 0.13$\pm$0.15 &  6.29$\pm$0.11 \\
$\rho$-correction HS$_{\rm{corr}}$ & 0.14$\pm$0.15&  6.21$\pm$0.11  \\
\hline
\end{tabular}
\caption{Mean and standard deviation of the $p_T$ signal reconstruction distributions for PYTHIA8 events embedded into $\rm{n_{PU}\!=\!60}$ minimum bias.}\label{table2}
\end{table}

\subsection{Heavy-ion collisions: impact of energy loss}\label{sec:hic}

\begin{figure}[ht]
\begin{center}
\includegraphics[scale=0.4]{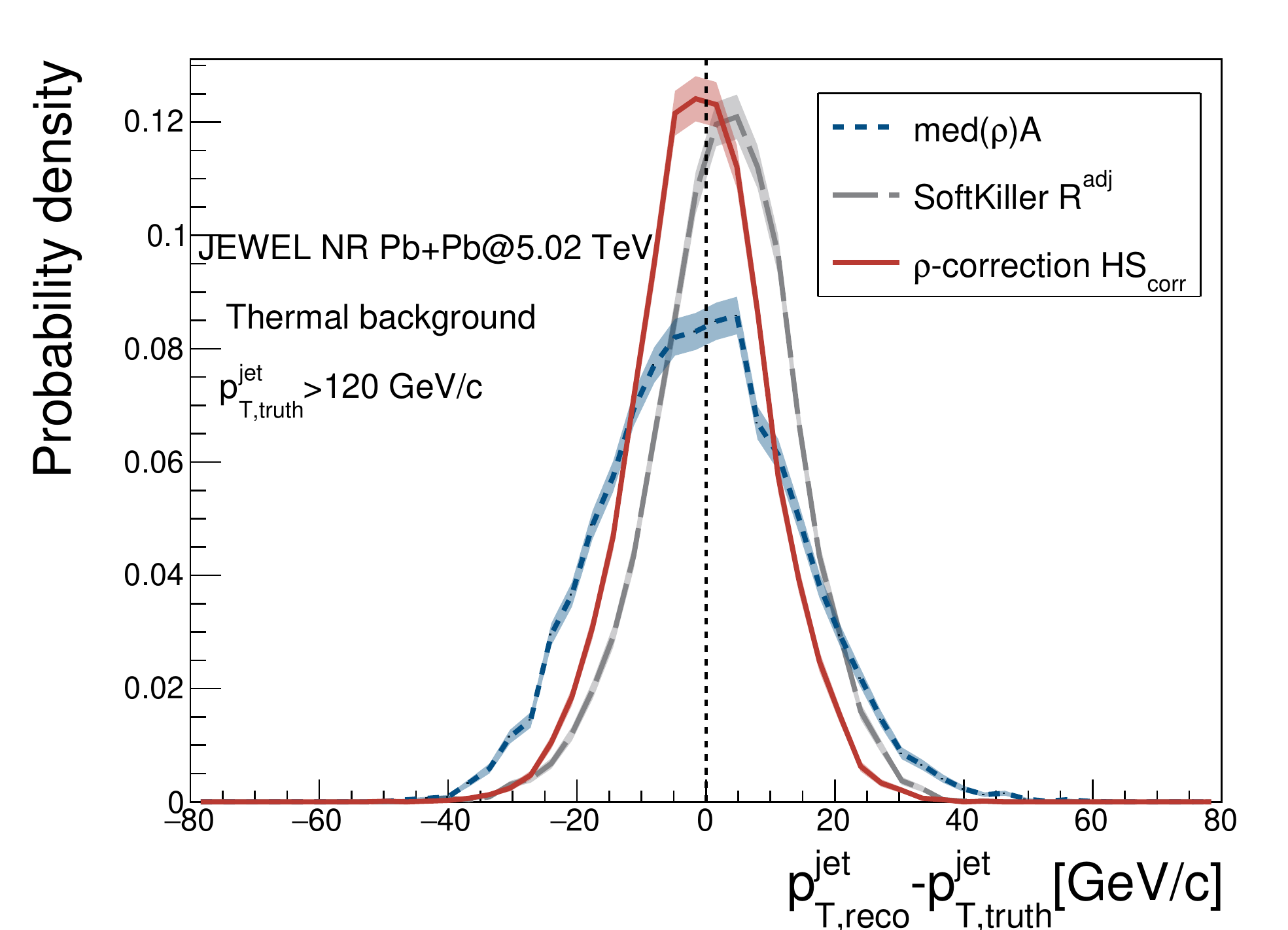}  
\end{center}
\vspace*{-0.5cm}
\caption[a]{Normalized $p_T$ signal reconstruction distributions for the median approach (dotted, blue), SoftKiller $R^{\rm{adj}}$ (dashed, gray) and $\rho$-correction HS$_{\rm{corr}}$ (solid, red). The colored error bands account for statistical fluctuations.}
\label{fig4}
\end{figure}

The application of the $\rho$-correction algorithm to a heavy-ion context starts by obtaining the value of $p_T^{\cut,{\rm{soft}}}$ by minimizing $\sigma(p_{T,\rm{reco}}^{\rm jet}\!-\!p_{T,\rm{truth}}^{\rm jet})$ within the JEWEL sample. The main source of uncertainty with respect to the p+p case is the impossibility of determining $p_{T,\rm{truth}}^{\rm jet}$ due to energy loss. Therefore, to avoid Monte Carlo dependence, the adjustment of free parameters in any method has to be done with p+p data. The standard deviation in vacuum is minimized, as shown in Fig.~\ref{fig3}, for a value of $p_T^{\cut,{\rm soft}}\!\sim\!2.5\mu$ for which we compute $p_{T,<}^{\rm{shift}}$ (both in the inclusive and the width-dependent case) in JEWEL without energy loss. In the case of SoftKiller $R^{\rm{adj}}$ we follow a similar procedure: first we optimize the value of $R_{kT}$ in JEWEL without energy loss $R_{kT}\!=\!0.21$ ($\langle p_{T}^{\cut}\rangle\!=\!3$, $\sigma(p_{T}^{\cut})\!=\!0.05$~GeV/c) and then use it in the heavy-ion context. 

These are all the ingredients that we need to apply Eq.~\ref{ourCut}. The two first statistical moments of the momentum reconstruction distributions in the JEWEL NR case are displayed in Table~\ref{table3}. Regarding the standard deviation we reach a similar conclusion to that in the p+p case: the $\rho$-correction HS$_{\rm corr}$ method reduces it by 0.6 GeV/c and 4.4 GeV/c when compared to SoftKiller $R^{\rm{adj}}$ and area-median, respectively.

The fact that the mean value within SoftKiller deviates from zero, as compared to the p+p case, reflects in-medium jet modification absent in the vacuum baseline. In the case of both $\rho$-correction and $\rho$-correction HS$_{\rm{corr}}$ the mean value of the $(p_{T,\rm{reco}}^{\rm jet}\!-\!p_{T,\rm{truth}}^{\rm jet})$-distribution is close to zero thanks to the $p_{T,<}^{\rm{shift}}$-term as given in Eqs.~\ref{shift_aa},\ref{shift_aa_hscorr}, respectively. This is a remarkable and non-trivial feature as the applicability of our unbiased method to heavy-ions doesn't lie on the common trade-off between resolution improvement and the shift on the mean due to energy-loss. The $\rho$-correction approach leads to simultaneously obtain a momentum reconstruction distribution centered at zero alike the area-median with a substantial reduction of its standard deviation along the lines of SoftKiller. 

For completeness, the momentum reconstruction distributions are shown in Fig.~\ref{fig4}. In there, the positive shift of SoftKiller together with the broadness of the area-median distribution are clearly visible. We conclude that $\rho$-correction HS$_{\rm{corr}}$ is an unbiased method ready to be used as background estimator in jet experimental analyses of heavy-ion collisions.

\begin{table}[ht]
\centering
\begin{tabular}{|c||c|c|}
\hline
 [GeV/c]&Mean& Standard deviation \\
\hline\hline
med($\rho$)A &-0.61$\pm$0.17 & 14.73$\pm$0.12 \\
SoftKiller $R^{{\rm adj}}$ & 2.90$\pm$0.12 & 10.91$\pm$ 0.09 \\
$\rho$-correction & 0.18$\pm$0.12 & 10.92$\pm$ 0.09  \\
$\rho$-correction HS$_{\rm{corr}}$ &-0.74$\pm$0.12 &10.34$\pm$ 0.08\\
\hline
\end{tabular}
\caption{Mean and standard deviation of the $p_T$ signal reconstruction distributions for JEWEL events embedded into the ($\langle N \rangle\!=\!7000$, $\mu\!=\!1.2$~GeV/c) background.}\label{table3}
\end{table}

\section{Summary and outlook} \label{Section5}
In this work we propose a new background estimation method dubbed $\rho$-correction that aims at improving the resolution on jet $p_T$ reconstruction analyses in hadronic collisions. To that end, we analyze the role of key elements of existing algorithms, such as the area-median and SoftKiller, in terms of their impact on the mean and standard deviation of the ($p_{T,\rm{reco}}^{\rm jet}\!-\!p_{T,\rm{truth}}^{\rm jet}$)-distribution. In the spirit of SoftKiller, we introduce a $p_T^{\cut}$ at the constituent level that we determine by minimizing the standard deviation of the jet momentum reconstruction distribution, in contrast to their mean value optimization. The price to pay for having a reduction on the standard deviation thanks to a $p_T^{\cut}$ is an overestimation of the background component due to abundant soft QCD radiation. Our method, summarized in a compact way by Eq.~\ref{ourCut}, overcomes this offset with an experimental accessible quantity: the average momentum stored below a given $p_T^\cut$ inside a jet to be measured in a clean environment like low pileup p+p collisions. We show how the average of this distribution is enough to correct for the shift in high-luminosity proton-proton collisions while a couple of additional measurable terms are needed in a heavy-ion context due to in-medium jet modification. The method results into a comparable performance to SoftKiller in the high-luminosity scenario while it outperforms both area-median and SoftKiller in heavy-ion collisions.  

Finally, we show that exploiting correlations between the soft and hard sectors within a jet, signal fluctuations are mitigated and, consequently, the jet momentum reconstruction resolution is improved with respect to the area-median and SoftKiller methods. The use of QCD correlations generated in the branching process of the shower to reduce the resolution in jet momentum reconstruction is the main novelty of this work. The natural continuation would be to implement a machine learning set up, in the spirit of \cite{Haake:2018hqn}, to systematically identify the set of substructure observables that better correlate with $\langle p_{T,<}^{\rm{sig}}\rangle$. Work in this direction is on-going. Finally, the combination of this background estimator with a novel subtraction method will be the subject of a follow-up publication~\cite{sharedLayer}.

\section*{ACKNOWLEDGMENTS} 
We would like to express our gratitude to Antonio Bueno, Megan Connors, Kolja Kauder and Brian Page for helpful discussions during the realization of this work. We thank Gavin Salam for a careful reading of the manuscript and sharing his insights into the SoftKiller method. Y. M. T. and A. S. O.'s work was supported by the U.S. Department of Energy, Office of Science, Office of Nuclear Physics, under contract No. DE- SC0012704,
and by Laboratory Directed Research and Development (LDRD) funds from Brookhaven Science Associates.
\bibliography{references_YamCut}{}
\bibliographystyle{apsrev4-1}

\end{document}